\documentclass{article}
\usepackage{amsfonts, amsmath, amssymb, amsthm, amstext}
\usepackage{graphicx, float}
\usepackage{siunitx}
\usepackage{braket}
\usepackage{authblk}
\usepackage{titlesec}
\usepackage[title]{appendix}
\usepackage{dsfont}
\usepackage{bm}
\usepackage{tikz}
\usepackage{cuted}
\usepackage{subcaption}
\usepackage{etoolbox}
\usetikzlibrary{arrows.meta,positioning,shapes.geometric}
\usepackage[english]{babel}
\usepackage[autostyle, english = american]{csquotes}
\MakeOuterQuote{"}
\usepackage[backend=biber,style=numeric,sorting=none]{biblatex}
\usepackage[colorlinks=true,linkcolor=red,citecolor=blue,urlcolor=blue]{hyperref}
\usepackage{sectsty}
\usepackage{indentfirst}
\usepackage{float}
\usepackage[ruled,vlined,linesnumbered]{algorithm2e}
\makeatletter
\renewenvironment{abstract}
  {\par\vspace{-0.8\baselineskip}\small\noindent\ignorespaces}
  {\par\vspace{-0.6\baselineskip}}
\makeatother
\DontPrintSemicolon
\SetAlgoVlined
\SetKwProg{Fn}{function}{}{end}
\SetKwComment{Comment}{$\triangleright$~}{}
\SetCommentSty{emph}
\SetAlgoCaptionSeparator{.}
\SetKwInOut{Input}{Input}
\SetKwInOut{Output}{Output}
\SetKw{KwTo}{to}
\sectionfont{\centering\large\MakeUppercase}
\subsectionfont{\centering\normalsize}

\subsubsectionfont{\centering\smallfont\itshape}
\graphicspath{{images/}}
\usepackage[letterpaper, top=1in, bottom=1in, left=1in, right=1in, heightrounded]{geometry}

\title{\large\textbf{Sample Complexity for Embedded Multipartite Entanglement Witness via Pauli and Clifford Classical Shadows}}
\author{Ziran Zhang}
\affil{\small(Dated: December 30, 2025)}
\date{}

\begin{document}
\maketitle
\begin{abstract}
    \hspace{1.5em}Detecting multipartite entanglement in many qubit systems is measurement-intensive, motivating protocols that estimate only selected observables with provable efficiency. In this work we use the classical shadow protocol to study the sample complexity required to estimate a family of subsystem $n$-partite entanglement witness embedded in an larger $N$-qubit system. We derive ensemble dependent variance bounds that lead to qualitatively distinct scaling for the snapshots cost at fixed additive error $\epsilon$ with numerical simulations confirm these trends, exhibiting a clear crossover from Pauli favorable performance for local witness to Clifford favorable performance as the witness becomes more global. 
\end{abstract}

\section{Introduction}\label{Intro}
Learning an unknown quantum system is difficult due to its exponentially growth in state space and probabilistic nature. Although full quantum tomography can reconstruct the entire system but its computational cost becomes impractical as the number of independent parameters in a density matrix scales with the rank and dimension\cite{Haah_2017}. This motivates learning strategies that avoid full reconstruction of state when the goal is to predict only certain properties of $\rho$. Two representative directions are shadow tomography, which targets simultaneous estimation of many expectation values $\text{Tr}(O_i\rho)$ without reconstructing $\rho$, and neural-network quantum state tomograph which using variational methods to approximate the wavefunction with neural network\cite{aaronson2018shadowtomographyquantumstates,Torlai2018NNQST}.

In this work, we adopt the classical shadows framework\cite{Huang_2020} to estimate expectation values $\langle O\rangle=\text{Tr}(O\rho)$ for observables that are corresponding to a family of subsystem $n$-partite entanglement witness operator embedded in an $N$-qubit register. Instead of reconstructing the full quantum state, under the frame work of classical shadow, the computational cost is therefore governed by the statistical properties of the observable under the chosen measurement ensemble.Concretely, we study a system with fixed system size $N$ and a families of n-partite entanglement witness observable. We then benchmark how efficiently these witness can be estimated using the classical shadow under two measurement ensembles: local Pauli measurements and global Clifford measurements. 

In Sec.\ref{Intro} we review the relevant fundamental concepts needed through this work, including quantum state, measurement, and etc. In Sec.\ref{shadow} we introduce the classical shadow and the specific measurement ensembles we used in this study. Based on this two part, Sec.\ref{witness} presents our construction of the witness family $(W^{(N)}_n(\theta))$ and the numerical protocol used to evaluate its performance across subsystem with derivation of shadow norm bound and estimation of sampling complexity for Pauli and Clifford ensembles, and in the Sec.\ref{discussion} discussing and comparing these theoretical predictions with numerical simulation result.

\subsection{Separable state, product state, and GHZ state}
In general, a multipartite quantum system are based on a tensor product Hilbert space where the joint state of subsystems of subsystems prepared independently is the tensor product of their states\cite{nielsen00}. This leads to the ideas of product states and separable states.
A pure product is a vector state that factorizes across the bipartition:\begin{equation}
|\psi_{AB}\rangle=|\psi_A\rangle\otimes|\psi_B\rangle.
\end{equation}
This means subsystem $A$ and subsystem $B$ can be assigned independent pure states with no irreducible quantum correlations. To be more rigorous, using the Schmidt decomposition: any bipartite pure state can be written as\cite{nielsen00,10.1119/1.17904}\begin{equation}
    |\psi\rangle=\sum_i\lambda_i| i_A\rangle| i_B\rangle,\quad\lambda_i\geq0,\quad\sum_i\lambda_i^2=1,
\end{equation}
for orthonormal Schmidt bases $\{| i_A\rangle\}$ and $\{| i_B\rangle\}$, where $\lambda_i$ is the Schmidt rank. A key consequence is that $|\psi\rangle$ is product iff it has Schmidt rank of 1, equivalently iff the reduced state $\rho_A=\text{Tr}_B(|\psi\rangle\langle\psi|)$ (and $\rho_B$) is pure\cite{nielsen00}.

In practical experiment, we always deal with mixed states and a straightforward extension is the product mixed state:\begin{equation}
    \rho_{AB}=\rho_A\otimes\rho_B.
\end{equation}
No entanglement is more general than being a single tensor product state, and this motivates the definition of a  separable state\cite{Horodecki_2009}:$\rho_{AB}$ is separable iff it can be written as a convex combination of product states\begin{equation}
\label{separable}\rho_{AB}=\sum_{i=1}^kp_i\rho_A^{(i)}\otimes\rho_B^{(i)},\quad pi\geq0,\quad\sum_ip_i=1.
\end{equation}
Conceptually, Eq.\ref{separable} shows that a separable state is generated by a classical mixture over product state: a shared classical label $i$ is sample with probability $p_i$, and conditioned on $i$ the parties hold the product state $\rho_A^{(i)}\otimes\rho_B^{(i)}$. Thus any correlation arise from classical mixing, not from entanglement across the bipartition\cite{Horodecki_2009,Bravyi_2005}.

Moreover, Eq.\ref{separable} highlight an important feature of "no entanglement": the set of separable states forms a convex subset of the full state space. Consequently, any state $\rho$ that cannot be written in the form of Eq.\ref{separable} is called entangled. In practice, one can also characterize nonclassical correlations via Bell inequalities. A violation of Bell inequality certifies nonlocality and therefore implies entanglement\cite{PhysicsPhysiqueFizika.1.195}.

A paradigmatic family of multipartite entangled state is called the GreenBerger-Horne-Zeilinger states. For N-qubits. the canonical GHZ state is\cite{greenberger2007goingbellstheorem}:\begin{equation}
    \label{GHZ}
    | GHZ_N\rangle=\frac{\left(|0\rangle^{\otimes N}+|1\rangle^{\otimes N}\right)}{\sqrt{2}},
\end{equation}
since for any nontrivial bipartition $A| \bar{A}$ the reduced state is mixed. Explicitly, tracing out $\bar{A}$ yields
\begin{equation}
    \begin{aligned}
    \rho_A = \text{Tr}_{\bar{A}}\left(| \text{GHZ}_N \rangle \langle \text{GHZ}_N |\right) = \frac{1}{2} \left(|0\rangle \langle 0|\right)^{\otimes |A|} + \frac{1}{2} \left(|1\rangle \langle 1|\right)^{\otimes |A|},
\end{aligned}
\end{equation}
so $| GHZ_N\rangle$ cannot factorize across any cut and thus represents a genuinely $N$-partite entangled resource\cite{G_hne_2009}.
\begin{figure}[H]
    \centering
    \begin{subfigure}[b]{0.48\textwidth}
        \centering
        \includegraphics[width=\textwidth]{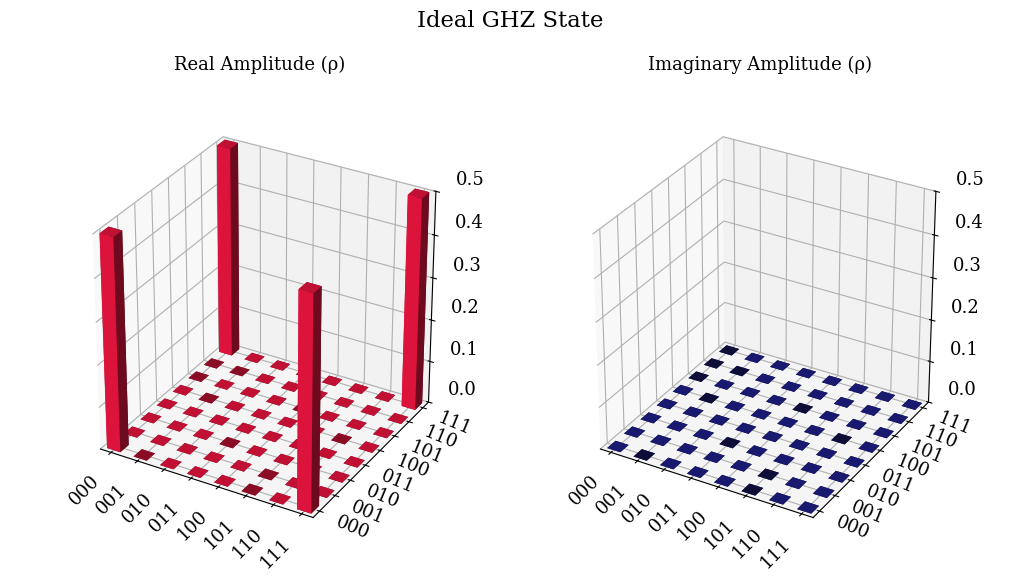}
        \caption{Ideal GHZ state in computational basis}
        \label{fig:idealGHZ}
    \end{subfigure}
    \hfill 
    \begin{subfigure}[b]{0.48\textwidth}
        \centering
        \includegraphics[width=\textwidth]{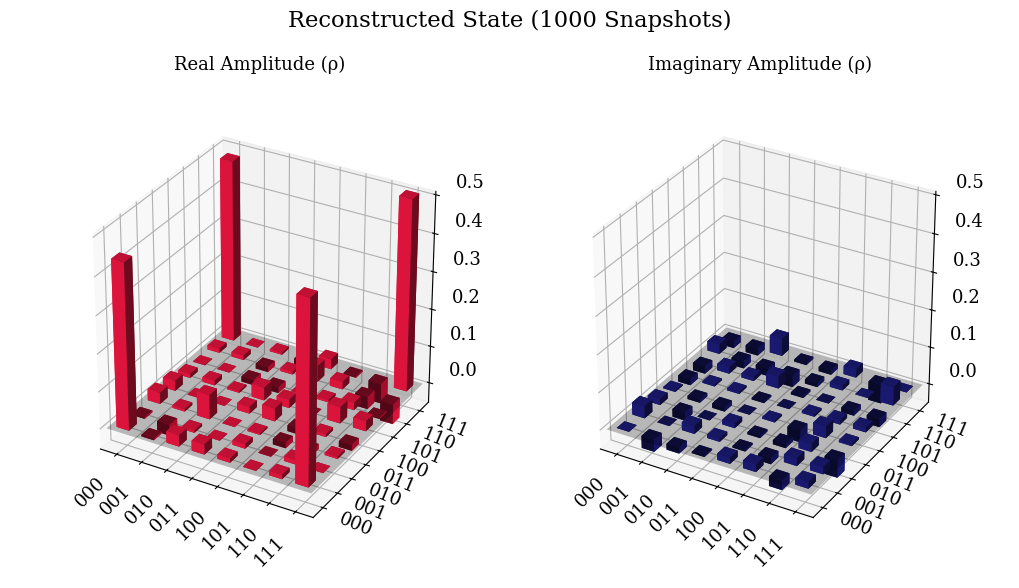}
        \caption{Reconstructed GHZ state via classical shadow}
        \label{fig:ReconstructGHZ}
    \end{subfigure}
    \caption{Comparison of the ideal and reconstructed GHZ states.}
    \label{fig:GHZ_Comparison}
\end{figure}
Fig.\ref{fig:GHZ_Comparison} compares the density matrix of an ideal 3-qubit GHZ state $\rho=|GHZ\rangle\langle GHZ|$ with the state reconstructed from classical shadow snapshots. For ideal GHZ state in panel (a), only the diagonal and the off-diagonal coherence are nonzero. Penal (b) presents the reconstructed density matrix. Despite the  presence of statistical noise (small fluctuations in the theoretically zero areas) resulting from the finite snapshots size, the reconstruction successfully resolve the four dominant corner peaks. This confirms that the classical shadow protocol captures both the classical outcome correlation and the phase coherence necessary to verify genuine multipartite entanglement, even with a limited snapshot budget.
\subsection{Classical Shadow protocol preview}
Before introduce the mathematical formalism of the classical shadow protocol, we briefly outline the physical picture of it. Instead of reconstructing the full state $\rho$, we repeatedly perform randomized measurements and collects the experimental data into a collection of snapshots $\{\hat{\rho_i}\}$ that can be reused to estimate different observables at later. Concretely, each snapshot consists of following steps\cite{Huang_2020}:\begin{enumerate}
    \item Randomize the measurement basis: Sample a random unitary $U_i\sim\mu$ from a chosen ensemble (in this work,  either local Pauli or global Clifford unitaries).
    \item  Measure in the computational basis: Apply $U_i$ to the unknown $N$-qubit sate $\rho$, then measure all qubits in the computation basis, obtaining a bit string outcome $b_i\in\{0,1\}^N$.
    \item Store the classical record $(U_i,b_i)$.
    \item Classical post processing into a snapshot: Map $(U_i,b_i)$ to an unbiased estimator(snapshot) $\hat{\rho}_i$.
\end{enumerate}
The detail description of the randomized measurement channel and the definition of an unbiased snapshot estimator are given in Appendix~\ref{app:measure}. The explicit explicit inversion maps for the ensembles in the work are provided in Appendix~\ref{app:ensembles}.

\section{Classical Shadow}\label{shadow}
The classical shadow protocol provides a method for predicting the expectation value of an arbitrary set of observables $O_1\dots O_M$ using a simple classical description of the quantum state $\rho$. Unlike full state shadow tomography, which required resource scaling exponentially with the system size $N$\cite{aaronson2018shadowtomographyquantumstates}, classical shadows allow for the prediction of $M$ properties with a properties with a sample complexity logarithmic in $M$\cite{Huang_2020}.
\subsection{Single shot estimation}
As we introduced in Sec.\ref{Intro} and Appendix~\ref{app:measure}, a single experimental shot yields a classical snapshot $\hat{\rho}=\mathcal{M}^{-1}(U^\dagger|\hat{b}\rangle\langle\hat{b}|)$. The expectation value of any observable $O$ can be estimated as following:\begin{equation}
    \hat{o}=\text{Tr}(O\hat{\rho})
\end{equation}
Since the snapshot is constructed to be the inverse of the measurement channel $\mathcal{M}$, this estimator is unbiased:\begin{equation}
    \begin{aligned}
        \mathbb{E}[\hat{o}]=\text{Tr}\left(O\mathbb{E}[\hat{\rho}]\right)=\text{Tr}\left(O\mathcal{M}^{-1}\left(\mathcal{M}\left(\rho\right)\right)\right)=\text{Tr}\left(O\rho\right).
    \end{aligned}
\end{equation}
However, the single shot estimator $\hat{\rho}$ is not necessarily a physical density matrix and can have large variance. The fluctuations of this estimator are governed by the shadow norm $|| O||^2_{shadow}$\cite{Huang_2020}:\begin{equation}
   \mathbb{E}_{U\sim\mathcal{U}}\left[\sum_b\langle b| U\sigma U^\dagger| b\rangle\langle b| U\mathcal{M}^{-1}(O_0)U^\dagger| b\rangle^2\right],
\end{equation}
where $O_0$ is the traceless part of the observable. The variance of a single shot estimated is bound by the norm $\text{Var}[\hat{o}]\leq|| O||^2_{shadow}$. This norm is critical as it quantifies the compatibility of an observable with the measurement ensemble $\mathcal{U}$.
\subsection{Sample complexity}
As stated by\cite{Huang_2020}, the classical shadow protocol allows for the simultaneous estimation of $M$ observables within additive error $\epsilon$ with failure probability $\delta$ where the number of measurement satisfies:\begin{equation}
    N_{tot}=\mathcal{O}\left(\frac{\log(M)}{\epsilon^2}\max_{1\leq i\leq M}|| O_i||^2_{\text{shadow}}\right).
\end{equation}
As we mentioned before that this sample complexity depends logarithmically on the number of observables $M$, effectively avoiding the exponentially growth with the dimension.
\subsection{Shadow norms for specific ensembles}
Global Clifford ensemble: For random global Clifford measurement, the shadow norm is bound ed by the Hilbert-Schmidt norm:\begin{equation}\label{cliff}
    ||O||^2_{\text{shadow}}\leq3\text{Tr}(O^2).
\end{equation}
This scaling is ideal for global observables, but it scales exponentially with the system size for local observables.

Local Pauli ensemble: For random local, the shadow norm depends on the locality of the observable rather than the system size. For a $k$ local observable, an operator acting non trivially on only $k$-qubits, the bound is:\begin{equation}\label{pauli}
    ||O||^2_{\text{shadow}}\leq4^k||O||^2_{\infty},
\end{equation}
Where quantity $||O||^2_{\text{shadow}}$ denotes the shadow norm of $O$ associated with the chosen measurement ensemble and $||O||_\infty$ for the operator norm.For observables that are tensor product of Pauli operators, this bound tightens to $3^k$\cite{Ippoliti_2024}. This exponential dependence on locality $k$ rather than system size $N$ makes the Pauli ensemble highly efficient for characterizing local properties in large many body system.

To make the above shadow norm bounds more concrete, we consider a specific example of a family of global state projector observables of the form\begin{equation}
    O(\theta)=|\psi(\theta)\rangle\langle\psi(\theta)|,
\end{equation}
where\begin{equation}\label{GHZeq}
    |\psi(\theta)\rangle=\cos\theta|0\rangle^{\otimes N}+\sin|1\rangle^{\otimes N},\quad\theta\in[0,\pi/4].
\end{equation}
Here $\theta$ directly controls the bipartite entanglement of $|\psi(\theta)$, since $\theta=0$ gives the product state $|0\rangle^{\otimes N}$, while $\theta=\pi/4$ gives the maximally entangled GHZ state $(|0\rangle^{\otimes N}+|1\rangle^{\otimes N})/\sqrt{2}$. 

\begin{figure}[H]
    \centering
    \includegraphics[width=0.75\linewidth]{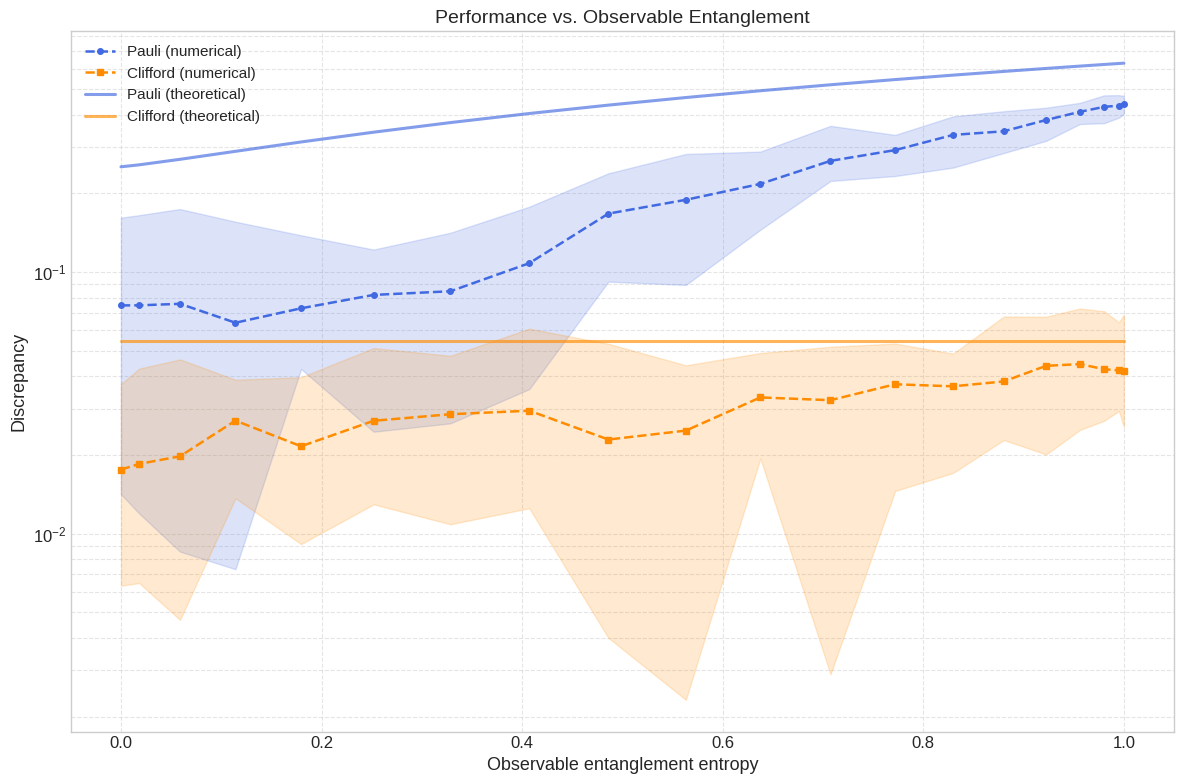}
    \caption{Theoretical and numerical discrepancy at fixed sanpshots versus observable entanglement entropy}
    \label{fig:entanglemententropy}
\end{figure}

In Fig.\ref{fig:entanglemententropy}, for each $\theta$ we fix a snapshot budget $N_\text{shots}=5000$ plotting the discrepancy (absolute estimation error)\begin{equation}
    \delta(\theta)=|\hat{o}(\theta) - o(\theta)|, \quad 
    \hat{o}(\theta)=\frac{1}{N_{\text{shots}}} \sum_{i=1}^{N_{\text{shots}}} \text{Tr}\left[ O(\theta) \hat{\rho_i} \right],
\end{equation}
as a function of the observable's entanglement entropy $S_A(\theta)$ for local Pauli and global Clifford shadows .Where $S_A(\theta)$ is defined as the von Neumann entanglement entropy of a fixed subsystem $A$ for the pure state underlying the observable. Concretely, by \cite{nielsen00} we have the definition of von Neumann entropy\begin{equation}
    S_A(\theta)=-\text{Tr}\left(\rho_A(\theta)\log_2\rho_A(\theta)\right)\quad\text{with}\,\rho_A(\theta)=\text{Tr}(|\psi(\theta)\rangle\langle\psi(\theta)|.
\end{equation}
In our simulation, we take $A$ to be the first qubit, so for the GHZ family(Eq.\ref{GHZeq}) we obtain the following form\begin{equation}
    S_A(\theta)=-\cos^2\theta\log_2(\cos^2\theta)-\sin^2\theta\log_2(\sin^2\theta).
\end{equation}

This trends is explained directly by Eqs.\ref{cliff} and \ref{pauli}. Under the global Clifford ensemble, the shadow norm is controlled by the Hilber-Schmidt norm $||O||^2_{\text{shadow}}\leq3\text{tr}(O^2)$. Since  $O(\theta)$ is a rank-1 projector, $\text{tr}(O)=1$, giving an $O(1)$ upper bound on the intrinsic estimator variance and therefore an error that remains approximately flat as the entanglement increases. In contrast, for local Pauli measurement the relevant control parameter is locality: $||O||^2_{\text{shadow}}\leq4^k||O||^2_{\infty}$. Here $||O(\theta)||_\infty=1$, but $O(\theta)$ is generically nonlocal. Consequently, at fixed $N_{shots}$, the Pauli shadow discrepancy increases with the observable entanglement.
\section{Generating Entanglement Witness}\label{witness}  
Entanglement witness is corresponding to a single Hermitian observable whose expectation value certifies entanglement. As we discussed before in Sec.\ref{Intro} the set of unentangled state is convex and closed, so any state outside this set can be separated from it. This separation picture leads to the standard definition: for a chosen reference set $B$, where $B$ is the chosen unentangled set (fully separable for bipartite entanglement, or biseparable for genuine multipartite entanglement), an operator $W=W^\dagger$ is an entanglement witness if \cite{Horodecki_2009,G_hne_2009}\begin{align}
    \text{Tr}(W\sigma)\geq0&\quad\text{for all}~\sigma\in B,\\
    \text{Tr}(W\rho)<0&\quad\text{for at least one}~\rho\notin B.
\end{align}
Consequently, for an experimental snapshot $\rho_\text{exp}$ an observed negative expectation value $\text{Tr}(W\rho_\text{exp})<0$ certifies that $\rho_\text{exp}\notin B$, or namely the $\rho_\text{exp}$ is entangled beyond the reference set $B$. By contrast, a nonnegative value $\text{Tr}(W\rho_\text{exp}\geq0$ is generally not sufficient to conclude $\rho_\text{exp}\in B$, but rather indicates that this particular witness doe not detect entanglement for the $\rho_\text{exp}$.

 Following\cite{Bourennane_2004}, a witness operator that detects genuine multipartite entanglement of a target pure state $\psi$ is given by \begin{equation}
    W(\psi)=\alpha\mathbb{I}-|\psi\rangle\langle\psi|,\quad\alpha=\max_{\phi \in B}|\langle\phi|\psi\rangle|^2,
\end{equation}
where $B$ is the reference set introduced above. Crucially, for the operator $W(\psi)$, the target must satisfy $|\psi\rangle\notin B$ (in particular, $|\psi\rangle$ mus be genuinely multipartite entangled when $B$ is chosen as the biseparable set), so that $\alpha<1$. Otherwise $\alpha=1$ and $W(\psi)$ cannot yield a negative expectation value. For the $N$-qubit GHZ state, a canonical choice is 
\begin{equation}
    W_{GHZ}=\frac{1}{2}\mathbb{I}-| GHZ_N\rangle\langle GHZ_N|,
\end{equation}
which detect genuine multipartite entanglement in a neighborhood of the GHZ state.
\subsection{Perturbed GHZ entanglement witness}
In practice, the experimentally prepared state may deviate from the ideal GHZ state, and a rigid GHZ projector can be suboptimal. To probe robustness and to explore a nearby family of witnesses, we introduce a perturbed target by coherently mixing the GHZ state with a separable state. Specifically, we first select a separable state $\rho_{\text{sep}}=|\phi\rangle\langle\phi|$ on the relevant $n$-qubit subsystem, where $|\phi\rangle$ is a normalized separable pure state (i.e., $|\phi\rangle\in B)$ chosen to lie close to the separable boundary of the unperturbed witness. Concretely, we require that it satisfies the witness constraint \begin{equation}
    \text{Tr}(\rho_{\text{sep}}W(\text{GHZ}_N))\approx 0.
\end{equation}
We then define the perturbed targe state 
    \begin{equation}  \label{perturebed state}      |\psi_{\text{pert}}(\theta)\rangle=\cos\theta| \text{GHZ}_N\rangle+\sin\theta|\phi\rangle,\quad0\leq\theta\leq1,
\end{equation}
and build the corresponding witness observable\begin{equation}
    W_{\text{pert}}(\theta)=\alpha(\theta)\mathbb{I}-|\psi_{\text{pert}}(\theta)\rangle\langle\psi_{\text{pert}}(\theta)|.
\end{equation}
To write this explicitly\begin{equation}
    \begin{aligned}
        |\psi_{\text{pert}}\rangle\langle\psi_{\text{pert}}|=\cos^2\theta|\text{GHZ}_n\rangle\langle\text{GHZ}_n|+\sin^2\theta|\phi\rangle\langle\phi|
        +\frac{1}{2}\sin(2\theta)\left(| \text{GHZ}_n\rangle\langle\phi|+|\phi\rangle\langle\text{GHZ}_n|\right),
    \end{aligned}
\end{equation}
Importantly, the $\alpha$ must be updated consistently with the perturbed direction:\begin{equation}
   \begin{aligned}\label{alphatheta}
        \alpha(\theta) = &\max_{\sigma \in B} \text{Tr}\left( |\psi_{\text{pert}}(\theta)\rangle \langle \psi_{\text{pert}}(\theta)| \sigma \right) = \max_{|\phi\rangle \in B} \left| \langle \phi | \psi_{\text{pert}}(\theta)\rangle \right|^2,
   \end{aligned}
\end{equation}
so $\alpha$ becomes $\theta$- dependent.

For the $n$-partite witness embedded in an $N$-qubit system, we enforce the support on the entire system by extending the projector with identity on the rest of the system:\begin{equation}\label{W_pert}
    \begin{aligned}
        &W_{n,\text{pert}}(\theta) = \alpha_n(\theta) \mathbb{I}_{2^N} - \left( \ket{\psi_{\text{pert}}(\theta)}\bra{\psi_{\text{pert}}(\theta)} \otimes \mathbb{I}_{2^{N-n}} \right),
    \end{aligned}
\end{equation}
which avoids turning the observable into a genuinely global operator through additional $|0\rangle\langle0|$ on the non entangled part of the system.For the expectation value $\hat{w}$ of this perturbed GHZ state witness, we compute the $\hat{w}$ per shot by\begin{equation}\label{witnessvalue}
    \hat{w}(n,\theta)=\text{Tr}(W_n(\theta)\hat{\rho}),
\end{equation}
so that $\mathbb{E}[\hat{w}(\theta)]=\text{Tr}(W_{\text{n,pert}}(\theta)\rho)$. The workflow for generating perturbed witness is then:
\begin{algorithm}
\caption{\textit{Perturbed GHZ-witness construction and shadow evaluation}.}
\label{alg:pertwitness}
\DontPrintSemicolon
\LinesNotNumbered
\SetKwComment{Comment}{$\triangleright$~}{}
\SetKwInOut{Input}{Input}
\SetKwInOut{Output}{Output}

\Input{$N$ total qubits; target block size $n$; perturbation angle $\theta$; witness scale $\alpha$; ensemble $e$ (Pauli/Clifford); shadow bank $S_e(\rho,S)=\{\hat\rho_1,\ldots,\hat\rho_S\}$.}
\Output{Shadow estimate $\widehat{w}(n,\theta)={\rm Tr}\big(W_{n,{\rm pert}}^{(N)}(\theta)\rho\big)$.}

\textbf{Step 1 (search separable state).}\;
Fix the unperturbed witness $W_n(0)=\alpha \mathbb{I}_{2^n}-\ket{{\rm GHZ}_n}\bra{{\rm GHZ}_n}$ and choose
$\rho_{\rm sep}=\ket{\phi}\bra{\phi}$ (product on the $n$-block) such that
\[
{\rm Tr}\left(\rho_{\rm sep}W_n(0)\right)\approx 0 .
\]

\textbf{Step 2 (perturbed target state).}\;
Construct
\[
\begin{aligned}
&\ket{\psi_{\rm pert}(\theta)}
=\frac{\cos\theta\ket{{\rm GHZ}_n}+\sin\theta\ket{\phi}}
{\big\|\cos\theta\ket{{\rm GHZ}_n}+\sin\theta\ket{\phi}\big\|},
\qquad\\
&\rho_{\rm pert}(\theta)=\ket{\psi_{\rm pert}(\theta)}\bra{\psi_{\rm pert}(\theta)} .
\end{aligned}
\]

\textbf{Step 3 (perturbed witness observable).}\;
Define
\[
W_{n,{\rm pert}}^{(N)}(\theta)
=\alpha \mathbb{I}_{2^N}-\left(\rho_{\rm pert}(\theta)\otimes \mathbb{I}_{2^{N-n}}\right),
\]

\textbf{Step 4 (classical-shadow evaluation).}\;
Estimate the witness value by classical shadow
\end{algorithm}
\subsection{Clifford\&Pauli measurement Performance analysis }
\begin{figure}[H]
    \centering
    \includegraphics[width=0.9\linewidth]{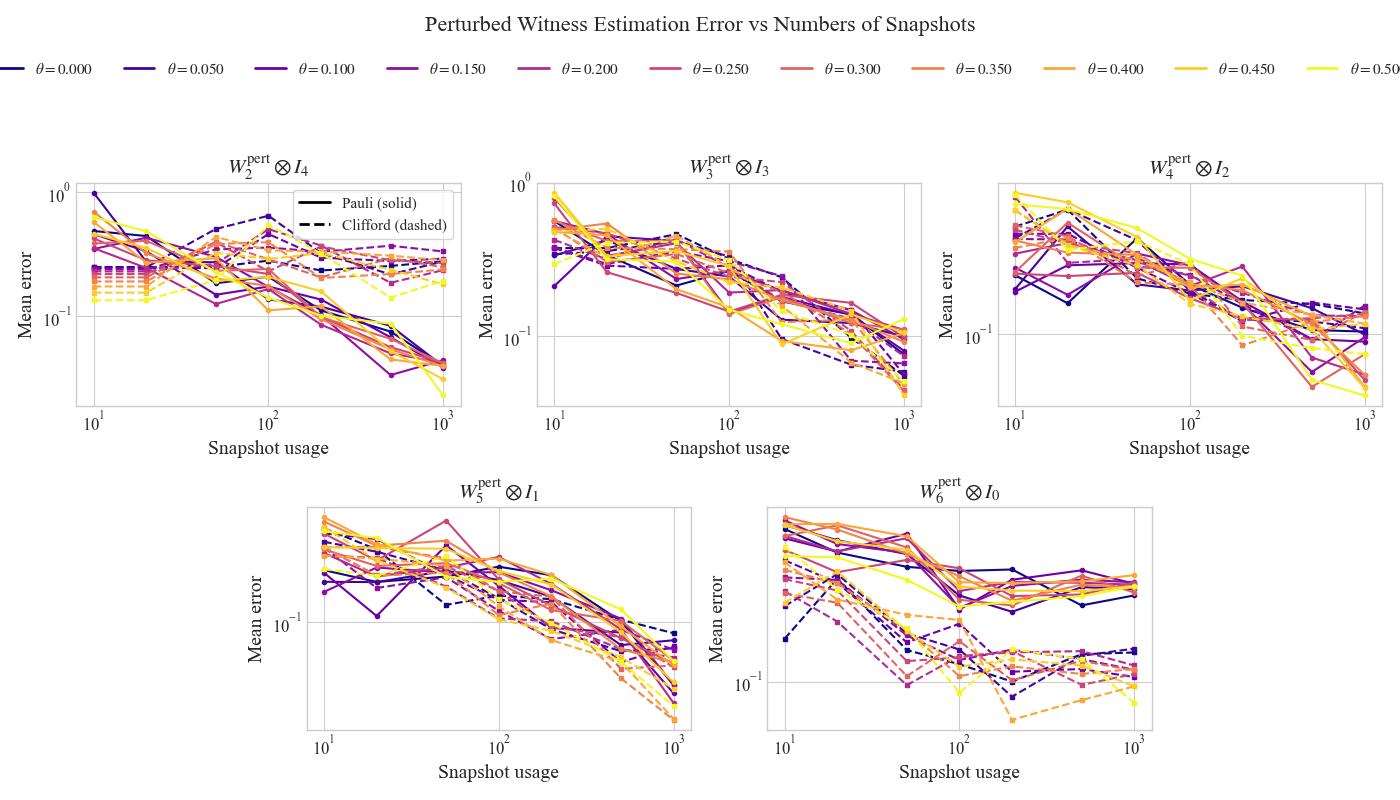}
    \caption{Estimation error $|\hat{w}-w_{\text{true}}|$ of Clifford and Pauli shadow at difference snapshot usage }
    \label{fig:errorshot}
\end{figure}

We start by comparing Pauli and Clifford classical shadows for estimating the perturbed witness expectation\begin{equation}
    W(n,\theta)=\text{Tr}(W^{(N)}_{n,\text{pert}}(\theta))
    \end{equation}
with $\rho_{\text{pert}}$ constructed from the GHZ state and the chosen separable state. For each $(n,\theta)$, Fig.\ref{fig:errorshot} presents the analysis of the estimation error $|\hat{w}-w_{\text{true}}|$ for the perturbed witness expectation value as a function of the snapshot usage. The results are stratified by the size of the entangled subsystem. For a 6-qubits system, the entangled subsystem ranging from a local 2-qubits witness $(W_w^{\text{pert}}\otimes \mathbb{I}_4)$ to the fully global 6-qubits witness $W_6^{\text{pert}}\otimes \mathbb{I}_0$.

Across all subsystem sizes, both measurement ensembles demonstrate the convergence behavior, but a noticeable difference appears at $W_w^{\text{pert}}\otimes I_4$ and $W_6^{\text{pert}}\otimes I_0$. This endpoint behavior is related to the embedded structure of $W^{(N)}_{n,\text{pert}}$ by Eq.\ref{W_pert}, where the statistical fluctuation of the estimator are dominated by the projector term $\rho_{\text{pert}}\otimes I_{2^{N-n}}$. In our implementation, the Clifford shadow performs a global channel inversion on the full $N$-qubits, while the Pauli shadow factorizes the inversion into a tensor product of single qubit inverse maps. As a  result, the variance responds differently to the $I_{2^{N-n}}$.

Classical shadow bounds show that the estimation error is controlled by the intrinsic variance of $X$\cite{Huang_2020}\begin{equation}
    X=\text{Tr}[(W_n^{pert}\left(\theta)\otimes \mathbb{I}_{2^{N-n}}\right)\hat{\rho}],
\end{equation}
which is bounded by the relevant shadow norm of the observable. For the global Clifford ensemble, this shadow nor admits a simple bound in terms of the traceless part of the observable. In our setting, the fluctuating contribution is dominated by the embedded projector term $O(\theta)=\rho_{\text{pert}}(\theta)\otimes I_{2^{N-n}}$, for which\begin{equation}
    \text{Tr}(O(\theta)^2)=\text{Tr}\left(O\left(\theta\right)\right)=\text{Tr}\left(\rho_{\text{pert}}\left(\theta\right)\right)\text{Tr}(I_{2^{N-n}})=2^{N-n}.
\end{equation}
Therefore, for the $N=6$, the most local case $n=2$ incurs the maximal penalty $2^{N-n}=2^4=16,$ explaining why the Clifford curves at $W_2^\text{pert}\otimes I_4$ remains higher error and converge more slowly with snapshot usage. By contrast, this penalty vanished when $n=N$, consistent with the comparatively improved Clifford performance for $W_6^\text{pert}\otimes I_0$.

In contrast, the Pauli performance becomes most inefficient in fully global witness, and is directly tied to the locality structure of the Pauli shadow. In our implementation. each Pauli snapshot is reconstructed qubit by qubit  via a tensor inverse channel\begin{equation}
    \hat{\rho}=\bigotimes^N_{i=1}\left(3U_i^\dagger|b_i\rangle\langle b_i|U_i-\mathbb{I}\right).
\end{equation}
This local reconstruction map explain the opposite trend, and as $n\xrightarrow[]{}N$, this locality advantage disappears and the dominant cost is instead set by the difficultly of resolving the high weight correlations in the projector component $\rho_{\text{pert}}(\theta)$.
\subsection{Shadow norm bound \& Sample complexity estimation}
The witness family used through this work, namely an $n$ entangled partite embedded into an $N$-qubits system,\begin{equation}
    W_{n,{\rm pert}}^{(N)}(\theta)
=\alpha \mathbb{I}_{2^N}-\big(\rho_{\rm pert}(\theta)\otimes \mathbb{I}_{2^{N-n}}\big).
\end{equation}
Since classical shadow snapshots have unit trace, the constant shift term $\alpha \mathbb{I}$ does not have contribution to the statistical fluctuation. and the variance is governed by the projector term $O(\theta)=\rho_{\text{pert}}(\theta)\otimes I_{2^{N-n}}$. Hence it is sufficed to study $||O(\theta)||^2$ for variance bound of measurement ensemble for this witness family\begin{equation}
    \text {Var}[\hat{w}]\leq||O(\theta)||^2_{\text{shadow}}.
\end{equation}
\subsection{Local Pauli measurement}
Under local Pauli shadows, the shadow norm is governed strictly by the locality of the observable. Where the relevant generic bound for a $k$-local observable $O$ is \begin{equation}
    ||O||^2_{\text{shadow, Pauli}}\leq4^k||O||^2_\infty,
\end{equation}
where $||O||_\infty$ is the operator norm\cite{Huang_2020}.
In our case, $O(\theta)$ acts non-trivially only on the $n$-qubits block, hence it is $k=n$ local. Thus we explicitly  evaluate the norm of the $O(\theta)$ by the multiplicative property of the operator norm\begin{equation}
    ||O(\theta)||_\infty=||\rho^{(n)}_{\text{pert}}(\theta)||_\infty\cdot||\mathbb{I}_{{2^{N-n}}}||_\infty=1\cdot1=1\quad\text{(rank-1 projector)}.
\end{equation}
Plugging into the locality bound gives \begin{equation}\label{witnessPauli}
    ||\rho_{\text{pert}}(\theta)\otimes \mathbb{I}_{2^{N-n}}
||_{\text{shadow, Pauli}}^2\leq4^n.
\end{equation}
This makes the Pauli shadows cost depend exponentially on multipartite size $n$, not on the full system size $N$. To estimate $\text{Tr}(W_n^{(N)}(\theta)\rho)$ to additive error $\epsilon$ over a set of $M$ candidate parameters $\{\theta_1,\dots\theta_M\}$. The classical shadow guarantee that the total number of snapshots can be chosen as \begin{equation}
    N_{\text{tot}}=\mathcal{O}\left(\frac{\log(M)}{\epsilon^2}\max_{1\leq i\leq M}||(W^{(N)}_{n,\rm pert}(\theta_i))||^2_\text{shadow}\right)=\mathcal{O}\left(\frac{4^n\log(M)}{\epsilon^2}\right)
\end{equation}
\subsection{Global Clifford measurement}
For the global Clifford shadows, the shadow norm bound controlled by the Hilbert-Schmidt norm. As derived in \cite{Huang_2020}, for any observable $O$ the general bound is:\begin{equation}
    ||O||^2_{\text{shadow}}\leq3\text{Tr}\left(O^2\right).
\end{equation}
For our witness observable, $W_{n,{\rm pert}}^{(N)}(\theta)$, we explicitly compute the trace of the squared observable \begin{equation}
    \text{Tr}(O(\theta)^2)=\text{Tr}(O(\theta))=\text{Tr}(\rho_{\text{pert}})\text{Tr}(\mathbb{I}_{2^{N-n}})=1\cdot2^{N-n}=2^{N-n}.
\end{equation}
Therefore\begin{equation}\label{witnessClifford}
    ||O(\theta)||^2_{\text{shadow,Clifford}}\leq3\cdot2^{N-n}.
\end{equation}
So the total number of snapshots needed for an additive error is\begin{equation}
    N_{\text{tot}}=\mathcal{O}\left(\frac{\log (M)}{\epsilon^2}3\cdot2^{N-n}\right).
\end{equation}
Thus, Clifford shadow cost increases primarily with the number of idle qubits $N-n$ in the observable $W_{n,{\rm pert}}^{(N)}(\theta)$, and decreases as the witness becomes more global as $n\rightarrow N$.

\section{Discussion}\label{discussion}
For a fixed error $(\epsilon=0.01)$, to validate Eq.\ref{witnessPauli} and Eq.\ref{witnessClifford}, we numerically simulated the estimation process for the snapshot cost required to estimate the perturbed witness expectation value to a fixed additive error $\epsilon=0.01$. Our derivation predicts qualitatively different scaling for the two ensembles. Consequently, at fixed $\epsilon$, we expect Pauli shadows to become more inefficient as the witness becomes more global, whereas Clifford snapshots become efficient as the number of idle qubit decreases.

For each configuration $(n,\theta)$, and each ensemble, we generate a bank of snapshots. Each snapshot produces a single shot unbiased estimator $\hat{w}_i(n,\theta)$, 
and the empirical single shot variance is computed as\begin{equation}
    \text{Var}[\hat{w}(n,\theta)]=
    \frac{1}{S-1}\sum^S_{I=1}(\hat{w}_i-\bar{w})^2,\quad \bar{w}=\frac{1}{S}\sum^S_{i=1}\hat{w_i},
\end{equation}
Where $S$ is the usage of snapshots. Since the variance of the sample mean scales as $\text{Var}(\bar{w}_S)=\text{Var}(\hat{w})/S$, the snapshot $S_{\text{req}}$ required to achieve $\epsilon$ is estimated by \cite{Huang_2020}\begin{equation}
    S_{\text{req}}\approx\frac{\text{Var}[\hat{w}(n,\theta)]}{\epsilon^2}
\end{equation}

\begin{figure}[H]
    \centering
    \includegraphics[width=1\linewidth]{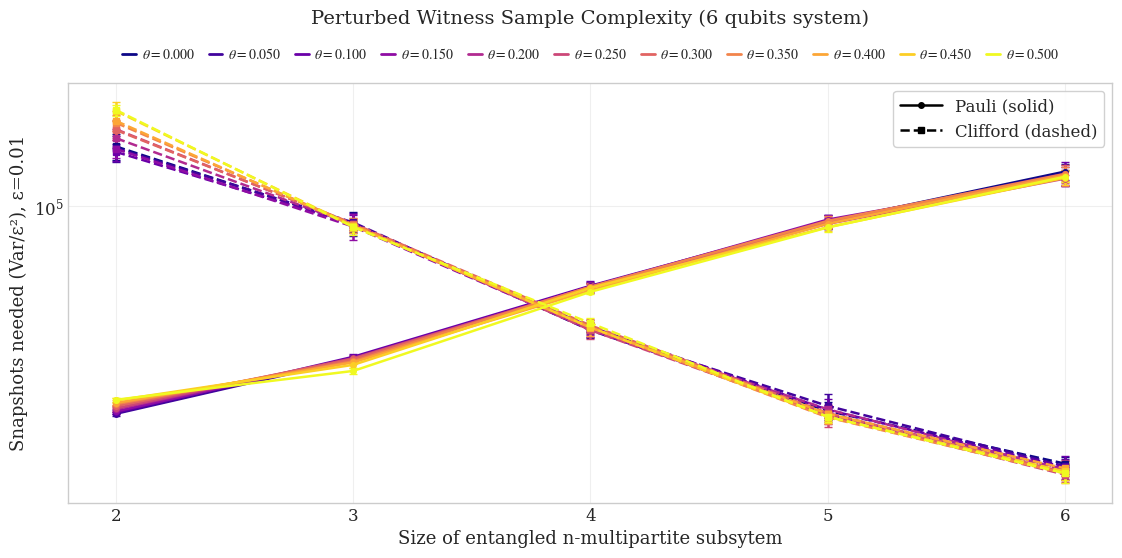}
    \caption{Numerical simulation result of 6 qubits system}
    \label{fig:6qubits}
\end{figure}

\begin{figure}[H]
    \centering
    \includegraphics[width=1\linewidth]{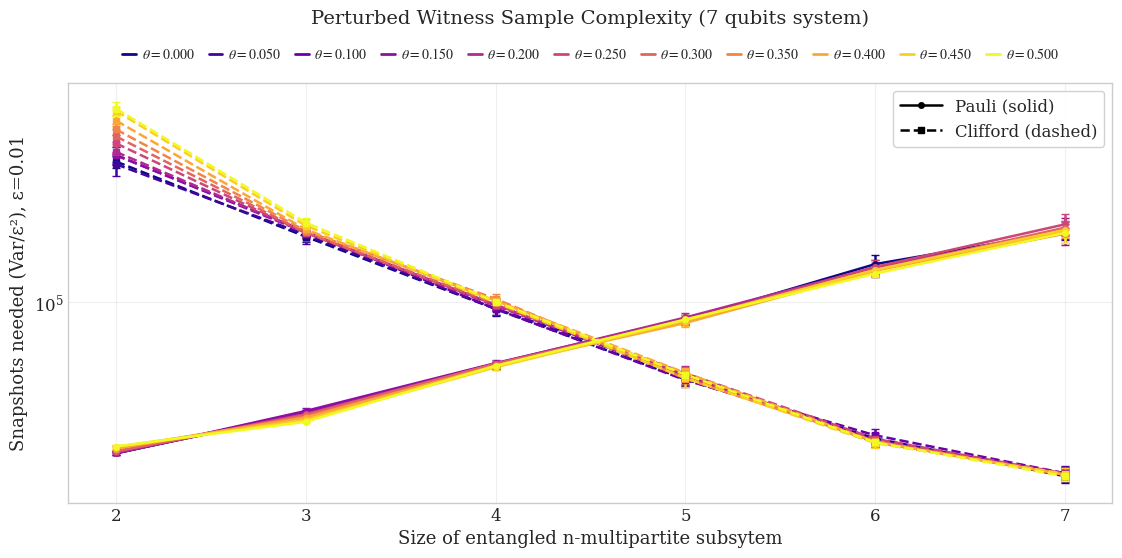}
    \caption{Numerical simulation result of 7 qubits system}
    \label{fig:7qubits}
\end{figure}

Fig.\ref{fig:6qubits} and Fig.\ref{fig:7qubits} demonstrate a clear and consistent agreement with the theoretical bounds derived in Eq.\ref{witnessPauli} and Eq.\ref{witnessClifford}. In both the $N=6$ and $N=7$ systems, the required snapshots for Pauli measurement increase monotonically with the entangled block size $n$ whereas the required snapshots for Clifford measurement decrease as $n$ grows.

Both plots show a crossover region where Pauli and Clifford require comparable snapshots. The reason is from comparing the bounds:\begin{equation}
    4^n\quad\text{vs} \quad2^{N-n}.
\end{equation}
As $n$ increases, Pauli worsens and Clifford improves, so they must cross once. The cross point also gives a clear performance boundary for $W_{n,{\rm pert}}^{(N)}(\theta)$, and at this point we can roughly proposed that for the $n$-multipartite in the size $N$ system if it satisfies $n\leq N/2$, the Pauli shadows become a more ideal choice.
\section{Conclusion}
We studied the sample complexity of estimating an $n$-partite entanglement witness family embedded in an $N$-qubit system using classical shadows, and compared local Pauli versus global Clifford measurement ensembles. The key observation is Pauli and Clifford shadows exhibit a crossover in shot cost: Pauli is more efficient for small, local witness with sample complexity growing roughly as $\sim4^n$, while Clifford becomes superior for more global witness with cost decreasing roughly as $\sim2^{N-n}$.

This crossover is directly explained by comparing the scaling $4^n$ versus $2^{N-n}$, and it provides a practical performance boundary for estimating multipartite entanglement witness in an $N$-qubit system, with a clear crossover around $n\approx N/2$: Pauli shadows are more efficient for witness supported on smaller $n$-qubit blocks, while Clifford shadows become superior as the witness approaches the system-wide $(n\rightarrow N)$ multipartite structure.
\section{Acknowledgments}
I would like to thank my instructor, Nadie Yiluo LiTenn, for guidance throughout this work and for helpful feedback on the development of the entanglement witness observable estimation analysis.
\newpage
\printbibliography

@article{Huang_2020,
   title={Predicting many properties of a quantum system from very few measurements},
   volume={16},
   ISSN={1745-2481},
   url={http://dx.doi.org/10.1038/s41567-020-0932-7},
   DOI={10.1038/s41567-020-0932-7},
   number={10},
   journal={Nature Physics},
   publisher={Springer Science and Business Media LLC},
   author={Huang, Hsin-Yuan and Kueng, Richard and Preskill, John},
   year={2020},
   month=jun, pages={1050–1057} }

@article{Bourennane_2004,
   title={Experimental Detection of Multipartite Entanglement using Witness Operators},
   volume={92},
   ISSN={1079-7114},
   url={http://dx.doi.org/10.1103/PhysRevLett.92.087902},
   DOI={10.1103/physrevlett.92.087902},
   number={8},
   journal={Physical Review Letters},
   publisher={American Physical Society (APS)},
   author={Bourennane, Mohamed and Eibl, Manfred and Kurtsiefer, Christian and Gaertner, Sascha and Weinfurter, Harald and Gühne, Otfried and Hyllus, Philipp and Bruß, Dagmar and Lewenstein, Maciej and Sanpera, Anna},
   year={2004},
   month=feb }

@article{Mele_2024,
   title={Introduction to Haar Measure Tools in Quantum Information: A Beginner's Tutorial},
   volume={8},
   ISSN={2521-327X},
   url={http://dx.doi.org/10.22331/q-2024-05-08-1340},
   DOI={10.22331/q-2024-05-08-1340},
   journal={Quantum},
   publisher={Verein zur Forderung des Open Access Publizierens in den Quantenwissenschaften},
   author={Mele, Antonio Anna},
   year={2024},
   month=may, pages={1340} }

@article{Horodecki_2009,
   title={Quantum entanglement},
   volume={81},
   ISSN={1539-0756},
   url={http://dx.doi.org/10.1103/RevModPhys.81.865},
   DOI={10.1103/revmodphys.81.865},
   number={2},
   journal={Reviews of Modern Physics},
   publisher={American Physical Society (APS)},
   author={Horodecki, Ryszard and Horodecki, Paweł and Horodecki, Michał and Horodecki, Karol},
   year={2009},
   month=jun, pages={865–942} }

@book{nielsen00,
  author = {Nielsen, Michael A. and Chuang, Isaac L.},
  publisher = {Cambridge University Press},
  title = {Quantum Computation and Quantum Information},
  year = 2000
}

@article{G_hne_2009,
   title={Entanglement detection},
   volume={474},
   ISSN={0370-1573},
   url={http://dx.doi.org/10.1016/j.physrep.2009.02.004},
   DOI={10.1016/j.physrep.2009.02.004},
   number={1–6},
   journal={Physics Reports},
   publisher={Elsevier BV},
   author={Gühne, Otfried and Tóth, Géza},
   year={2009},
   month=apr, pages={1–75} }

@article{Elben_2022,
   title={The randomized measurement toolbox},
   volume={5},
   ISSN={2522-5820},
   url={http://dx.doi.org/10.1038/s42254-022-00535-2},
   DOI={10.1038/s42254-022-00535-2},
   number={1},
   journal={Nature Reviews Physics},
   publisher={Springer Science and Business Media LLC},
   author={Elben, Andreas and Flammia, Steven T. and Huang, Hsin-Yuan and Kueng, Richard and Preskill, John and Vermersch, Benoît and Zoller, Peter},
   year={2022},
   month=dec, pages={9–24} }

@article{Elben_2019,
   title={Statistical correlations between locally randomized measurements: A toolbox for probing entanglement in many-body quantum states},
   volume={99},
   ISSN={2469-9934},
   url={http://dx.doi.org/10.1103/PhysRevA.99.052323},
   DOI={10.1103/physreva.99.052323},
   number={5},
   journal={Physical Review A},
   publisher={American Physical Society (APS)},
   author={Elben, A. and Vermersch, B. and Roos, C. F. and Zoller, P.},
   year={2019},
   month=may }

@article{Dankert_2009,
   title={Exact and approximate unitary 2-designs and their application to fidelity estimation},
   volume={80},
   ISSN={1094-1622},
   url={http://dx.doi.org/10.1103/PhysRevA.80.012304},
   DOI={10.1103/physreva.80.012304},
   number={1},
   journal={Physical Review A},
   publisher={American Physical Society (APS)},
   author={Dankert, Christoph and Cleve, Richard and Emerson, Joseph and Livine, Etera},
   year={2009},
   month=jul }

@article{Dehaene_2003,
   title={Clifford group, stabilizer states, and linear and quadratic operations over GF(2)},
   volume={68},
   ISSN={1094-1622},
   url={http://dx.doi.org/10.1103/PhysRevA.68.042318},
   DOI={10.1103/physreva.68.042318},
   number={4},
   journal={Physical Review A},
   publisher={American Physical Society (APS)},
   author={Dehaene, Jeroen and De Moor, Bart},
   year={2003},
   month=oct }

@misc{aaronson2018shadowtomographyquantumstates,
      title={Shadow Tomography of Quantum States}, 
      author={Scott Aaronson},
      year={2018},
      eprint={1711.01053},
      archivePrefix={arXiv},
      primaryClass={quant-ph},
      url={https://arxiv.org/abs/1711.01053}, 
}

@article{Ippoliti_2024,
   title={Classical shadows based on locally-entangled measurements},
   volume={8},
   ISSN={2521-327X},
   url={http://dx.doi.org/10.22331/q-2024-03-21-1293},
   DOI={10.22331/q-2024-03-21-1293},
   journal={Quantum},
   publisher={Verein zur Forderung des Open Access Publizierens in den Quantenwissenschaften},
   author={Ippoliti, Matteo},
   year={2024},
   month=mar, pages={1293} }

@article{Haah_2017,
   title={Sample-optimal tomography of quantum states},
   ISSN={1557-9654},
   url={http://dx.doi.org/10.1109/TIT.2017.2719044},
   DOI={10.1109/tit.2017.2719044},
   journal={IEEE Transactions on Information Theory},
   publisher={Institute of Electrical and Electronics Engineers (IEEE)},
   author={Haah, Jeongwan and Harrow, Aram W. and Ji, Zhengfeng and Wu, Xiaodi and Yu, Nengkun},
   year={2017},
   pages={1–1} }

@article{Flammia_2011,
   title={Direct Fidelity Estimation from Few Pauli Measurements},
   volume={106},
   ISSN={1079-7114},
   url={http://dx.doi.org/10.1103/PhysRevLett.106.230501},
   DOI={10.1103/physrevlett.106.230501},
   number={23},
   journal={Physical Review Letters},
   publisher={American Physical Society (APS)},
   author={Flammia, Steven T. and Liu, Yi-Kai},
   year={2011},
   month=jun }

@misc{greenberger2007goingbellstheorem,
      title={Going Beyond Bell's Theorem}, 
      author={Daniel M. Greenberger and Michael A. Horne and Anton Zeilinger},
      year={2007},
      eprint={0712.0921},
      archivePrefix={arXiv},
      primaryClass={quant-ph},
      url={https://arxiv.org/abs/0712.0921}, 
}

@misc{gottesman1998heisenbergrepresentationquantumcomputers,
      title={The Heisenberg Representation of Quantum Computers}, 
      author={Daniel Gottesman},
      year={1998},
      eprint={quant-ph/9807006},
      archivePrefix={arXiv},
      primaryClass={quant-ph},
      url={https://arxiv.org/abs/quant-ph/9807006}, 
}

@article{10.1119/1.17904,
    author = {Ekert, Artur and Knight, Peter L.},
    title = {Entangled quantum systems and the Schmidt decomposition},
    journal = {American Journal of Physics},
    volume = {63},
    number = {5},
    pages = {415-423},
    year = {1995},
    month = {05},
    issn = {0002-9505},
    doi = {10.1119/1.17904},
    url = {https://doi.org/10.1119/1.17904},
    eprint = {https://pubs.aip.org/aapt/ajp/article-pdf/63/5/415/11806926/415_1_online.pdf},
}

@article{Torlai2018NNQST,
  author  = {Torlai, Giacomo and Mazzola, Guglielmo and Carrasquilla, Juan and Troyer, Matthias and Melko, Roger and Carleo, Giuseppe},
  title   = {Neural-network quantum state tomography},
  journal = {Nature Physics},
  year    = {2018},
  volume  = {14},
  number  = {5},
  pages   = {447--450},
  doi     = {10.1038/s41567-018-0048-5},
  url     = {https://doi.org/10.1038/s41567-018-0048-5}
}

@article{Bravyi_2005,
   title={Universal quantum computation with ideal Clifford gates and noisy ancillas},
   volume={71},
   ISSN={1094-1622},
   url={http://dx.doi.org/10.1103/PhysRevA.71.022316},
   DOI={10.1103/physreva.71.022316},
   number={2},
   journal={Physical Review A},
   publisher={American Physical Society (APS)},
   author={Bravyi, Sergey and Kitaev, Alexei},
   year={2005},
   month=feb }

@article{PhysicsPhysiqueFizika.1.195,
  title = {On the Einstein Podolsky Rosen paradox},
  author = {Bell, J. S.},
  journal = {Physics Physique Fizika},
  volume = {1},
  issue = {3},
  pages = {195--200},
  numpages = {6},
  year = {1964},
  month = {Nov},
  publisher = {American Physical Society},
  doi = {10.1103/PhysicsPhysiqueFizika.1.195},
  url = {https://link.aps.org/doi/10.1103/PhysicsPhysiqueFizika.1.195}
}
\newpage
\begin{appendices}
\section{\texorpdfstring{\NoCaseChange{Randomized measurement and measurement channel}}{Randomized measurement and measurement channel}}
\label{app:measure}
In this appendix, we discuss the detail mathematical formalism of constructing measurement channel in classical shadow protocol.
We model each experimental "shot" as a randomized basis measurement on an unknown $N$-qubit state $\rho$. By sampling a random unitary $U$ from specified ensemble (local Pauli or global Clifford), apply it to the state, and then measure all qubits in the computational basis$\{| b\rangle\}_{b\in\{0,1\}^N}$, recording the bit string outcome $b$. Which corresponds to a projective measurement on the rotated state $U\rho U^\dagger$, and we may treat it as a POVM (Positive Operator Valued Measure) on $\rho$ with elements $E_b(U)=U^\dagger| b\rangle\langle b| U $ that satisfy:\begin{equation}
    \label{EbU}
    \sum_b E_b\left(U\right)=I,
\end{equation}
and give probabilities via the Born rule $\rho(b| U)=\text{Tr}(E_b(U)\rho)$\cite{Mele_2024}.

A standard way to implement random unitary measurement is to pick a subgroup $\mu$ of the full unitary group $U$ over the $N$ qubits. On each shot, sample $U\sim\mu$ and apply to an unknown state $\rho$, and then measure in the computational basis. The observed outcome $b$ is distributed according to Born's rule. Averaging over the random choice of $U\sim\mu$ and summing over all possible measurement outcomes $b$ defines a measurement channel $\mathcal{M}$ that maps states to the expected snapshots\begin{equation}\begin{aligned}
    \mathcal{M}(\rho) := \sum_{b=1}^{d} \mathbb{E}_{U \sim \mu} \left[ \langle b | U \rho U^{\dagger} | b \rangle U^{\dagger} | b \rangle \langle b | U \right] = \mathbb{E} \left[ U^{\dagger} | \hat{b} \rangle \langle \hat{b} | U \right],
\end{aligned}
\end{equation}
Where $\hat{b}$ denotes the actually observed outcome, and if $M$ is invertible one defines an unbiased snapshot\begin{equation}
    \hat{\rho}=\mathcal{M}^{-1}\left(U^\dagger|\hat{b}\rangle\langle\hat{b}| U\right).
\end{equation}
By construction, the snapshot is unbiased in the sense that averaging over the measurement randomness recovers the true state:$\mathbb{E}[\hat{\rho}]=\rho$, and then any expectation value $o=\text{Tr}(O\rho)$ can be estimated unbiasedly by $\hat{o}=\text{Tr}(O\hat{\rho})$\cite{Huang_2020}.
\section{\texorpdfstring{\NoCaseChange{Pauli and Clifford ensembles}}{Pauli and Clifford ensembles}}
\label{app:ensembles}
In this appendix, we define the measurement ensemble used throughout the paper and summarizing the corresponding measurement channels in the classical shadow formalism.
\subsection{Pauli measurement}
Pauli measurements correspond to choosing $U$ from the local Clifford ensemble,i.e. a tensor product of single qubit Clifford rotations\begin{equation}
U = \bigotimes_{j=1}^{N} U_j, \quad U_j \in \mathcal{C}_1,
\end{equation}
where each $U_j$ is drawn uniformly from the single qubit Clifford group $\mathcal{C}_1$.
Because single qubit Clifford maps the $Z$-basis to eigenbasis of $X$, $Y$, or $Z$, this is equivalent to measuring each qubit in a uniformly random Pauli basis${X,Y,Z}$. In group theoretic terms, the relevant operator basis is the $N$-qubit Pauli group\cite{gottesman1998heisenbergrepresentationquantumcomputers}\begin{equation}
    \mathcal{P}_N=\{P_1\otimes\dots\otimes P_N:P_j\in\{I,X,Y,Z\}\}.
\end{equation}
so one may treat each shot as picking a random Pauli string and the advantage is that Pauli measurements require only single qubit rotations and keeping circuit depth minimal\cite{Elben_2022,Elben_2019}.

In the perspective of classical shadow, choosing random single qubit Pauli measurement basis induces a measurement channel that factorizes across qubits. Specifically, on each shot we sample a measurement basis uniformly from $\{X, Y,Z\}$ which defines a single qubit channel $\mathcal{M}_1$, and the corresponding $N$-qubit channel $\mathcal{M}_N$is the tensor product\cite{Elben_2022}\begin{equation}
    \mathcal{M}_N=\mathcal{M}_1^{\otimes  N}.
\end{equation}
On the Pauli operator basis, $\mathcal{M}_1$ acts diagonally as
\begin{equation}
  \mathcal{M}_1\left(I\right) = I, \quad\mathcal{M}_1(P) = \frac{1}{3}P\quad\text{where}\; P \in \{X, Y, Z\},
\end{equation}
Consequently, for any Pauli string $P$ of weight $k$, the computational cost for estimating a Pauli observable scales as $3^k$\cite{Huang_2020}.
This locality makes Pauli measurements especially effective for low weight(local) observables, while performance can degrade for high weight(global) observable\cite{Flammia_2011}.
\subsection{Clifford measurement}
Clifford measurements replace the random unitary ensemble from local single qubit rotation to a global $N$-qubit Clifford $U\in\mathcal{C}_N$. The Clifford group is defined as the normalizer of the Pauli group\cite{gottesman1998heisenbergrepresentationquantumcomputers}:\begin{equation}
    \mathcal{C}_N=\{U\in U(2^N):U\mathcal{P}_NU^\dagger=\mathcal{P}_N\},
\end{equation}
so conjugation by any Clifford maps Pauli strings to Pauli strings. Moreover, sampling $U\sim U(\mathcal{C}_N)$ and measuring in the computational basis is equivalent to measuring $\rho$ in a random stabilizer basis, because\begin{equation}
    E_{U,b}=U^\dagger| b\rangle\langle b| U
\end{equation}
is a stabilizer state projector\cite{Dehaene_2003}. A key advantage is that the uniform distribution over the Clifford group forms a unitary 2-design meaning it reproduces Haar second moment statistics for all tasks depending only on second moment\cite{Dankert_2009}. This property implies $\mathcal{M}$ becomes an isotropic depolarizing map:\begin{equation}
   \begin{aligned}
        \mathcal{M}\left(\rho\right)=\sum_b\mathbb{E}_U[\text{Tr}\left(\rho E_{u,b}\right)E_{u,b}]=\frac{1}{d+1}\left(\text{Tr}\left(\rho\right)I+\rho\right)
   \end{aligned}
\end{equation}
and it can be inverted explicitly, yielding the single inverted snapshot for random Clifford basis.\begin{equation}
    \hat{\rho}=(2^n+1)U^\dagger|\hat{b}\rangle\langle\hat{b| U-I}.
\end{equation}
This closed form expression is one reason that global Clifford measurement are often efficient for estimating global properties (e.g. fidelity), because the measurement channel is isotropic rather than a tensor product of single qubit channels.
\end{appendices}

\end{document}